\documentclass[journal,preprint]{IEEEtran}

\ifCLASSINFOpdf
  \usepackage[pdftex]{graphicx}

\else

  \usepackage[dvips]{graphicx}

\fi

\begin{document}

\title{Three-dimensional simulations of laser-plasma interactions at ultrahigh intensities}

\author{Frederico Fi\'uza, 
        Ricardo A. Fonseca, Lu\'is O. Silva, 
        John Tonge, Joshua May, and Warren B. Mori
\thanks{F. Fiuza and L. O. Silva are with the GoLP/Instituto de Plasmas e Fus\~ao Nuclear - Laborat\'orio Associado, Instituto Superior T\'ecnico, 1049-001 Lisboa, Portugal (e-mail: frederico.fiuza@ist.utl.pt; luis.silva@ist.utl.pt).}
\thanks{R. A. Fonseca is with the GoLP/Instituto de Plasmas e Fus\~ao Nuclear - Laborat\'orio Associado, Instituto Superior T\'ecnico, 1049-001 Lisboa, Portugal and also with the Departamento Ci\^encias e Tecnologias da Informa\c c\~ao, Instituto Superior de Ci\^encias do Trabalho e da Empresa, 1649-026 Lisboa, Portugal.}
\thanks{J. Tonge, J. May, and W. B. Mori are with the University of California at Los Angeles, Los Angeles, CA 90095 USA.}
\thanks{This work was supported by Funda\c c\~ao para a Ci\^encia e Tecnologia (Portugal) through the Grants PTDC/FIS/66823/2006 and SFRH/BD/38952/2007, by PRACE through the project PRA030, by the European Community (HiPER project EC FP7 no. 211737), by the NSF under Grant Nos. PHY-0078508 and PHY-0904039, and by the DOE under Contract Nos. DE-FG03-09NA22569, and under the Fusion Science Center for Matter Under Extreme Conditions.}}

\maketitle

\begin{abstract}
Three-dimensional (3D) particle-in-cell (PIC) simulations are used to investigate the interaction of ultrahigh intensity lasers ($> 10^{20}$ W/cm$^{-2}$) with matter at overcritical densities. Intense laser pulses are shown to penetrate up to relativistic critical density levels and to be strongly self-focused during this process. The heat flux of the accelerated electrons is observed to have an annular structure when the laser is tightly focused, showing that a large fraction of fast electrons is accelerated at an angle. These results shed light into the multi-dimensional effects present in laser-plasma interactions of relevance to fast ignition of fusion targets and laser-driven ion acceleration in plasmas. 
\end{abstract}


\IEEEpeerreviewmaketitle

\IEEEPARstart{T}{he} interaction of ultraintense laser pulses with matter has opened the way to the exploration of highly nonlinear physical regimes of interest for many applications such as fast ignition of fusion targets \cite{bib:tabak,bib:kodama} or compact plasma based accelerators \cite{bib:dawson,bib:malka}. In many of these scenarios, the laser frequency is lower than the plasma frequency of the ionized target (overcritical target) and therefore the laser cannot penetrate deep into the plasma, being reflected and/or absorbed. 

The detailed study of laser absorption is crucial in order to understand the generation of fast electrons in these regimes. However, overcritical targets are difficult to probe experimentally and therefore many experimental results rely on particle-in-cell (PIC) simulations in order to better understand laser absorption and fast electron acceleration. Most of the simulation studies are limited to one-dimensional (1D) \cite{bib:kemp} and two-dimensional (2D) \cite{bib:tonge} simulations, due to the need to resolve the fine temporal and spacial scales of plasmas at very high densities.

In this paper we present 3D OSIRIS \cite{bib:osiris} simulations of the interaction of ultraintense laser pulses with overcritical targets in order to study some of the multi-dimensional features of particle acceleration/laser absorption in these regimes. We use a laser pulse with an intensity of $2\times10^{20}$ W/cm$^{-2}$ (normalized vector potential $a_0 = 12$), central wavelength $\lambda_0 = 1 ~\mu$m, focused to a spot size of 5 $\mu$m at the target front. The pulse has a gaussian transverse profile and a flat-top temporal profile with 60 fs gaussian rise time. The target consists of a deuterium plasma with a density gradient going from the critical density, $n_c$, to 50 $n_c$, scale length $L_g = 2.5 ~\mu$m, followed by a 30 $\mu$m region at $50~ n_c$.  The simulation box has a size of $50 \times 30 \times 30$ $\mu$m$^3$, which is resolved with $1416\times848\times848$ cells, and is run up to 500 fs. Each cell has 8 electrons and 8 ions for a total of $\sim$ 15 billion particles.

\begin{figure*}[!t]
\centering
\includegraphics[width=7.1in]{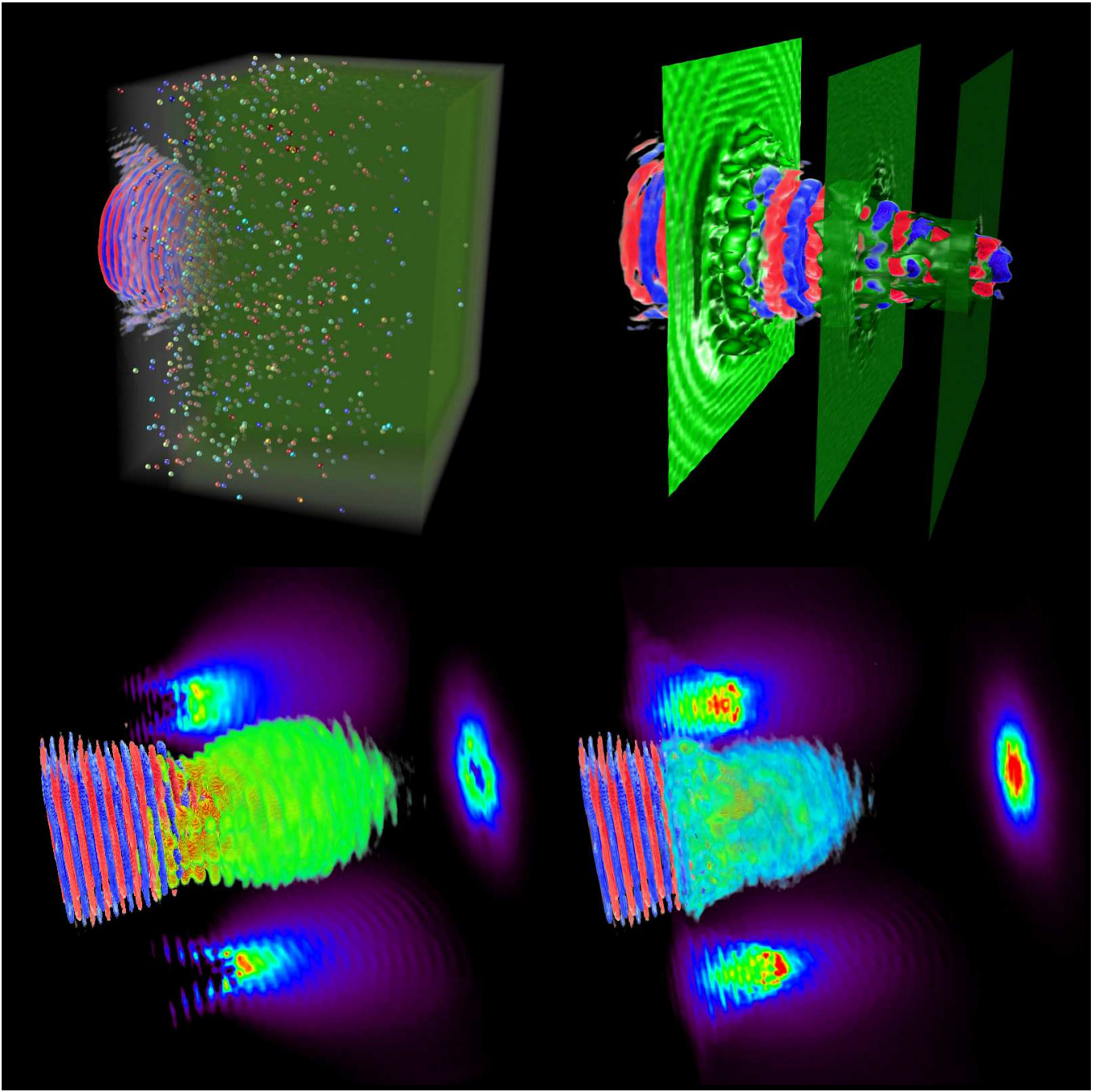}
\caption{Interaction of an ultrahigh intensity laser pulse with an overcritical plasma slab. (top-left) The laser pulse (red-blue) hits a deuterium plasma (green) from the left-hand-side of the simulation box, accelerating electrons (spheres colored by energy from 0 to 15 MeV: blue-yellow-red). (top-right) Laser (red-blue) penetration and self-focusing in the overcritical plasma. The different isosurfaces (green) correspond to different plasma densities: $n_c$ (left plane), 5 $n_c$ (middle plane), and 15 $n_c$ (right plane). (bottom-left) Electron heat-flux (violet-blue-green-red) generated by the strong laser (red-blue). The projections correspond to the heat flux in the different simulation planes, illustrating the generation of an annular pattern in the laser propagation direction. (bottom-right) Plasma electron kinetic energy density (violet-blue-green-red) when hit by the laser pulse (red-blue). The projections correspond to the kinetic energy density in the different simulation planes.} 
\label{fig:fig_3D}
\end{figure*}

Fig. \ref{fig:fig_3D} illustrates the interaction of the laser pulse with the overcritical target after 400 fs of propagation as well as the main features of the fast electron population that is generated during the interaction. The laser pulse is observed to penetrate the target up to $> 15 ~n_c$ due to relativistic effects and to be continuously self-focused during this process down to a spot size of  $\sim 1 ~\mu$m. This penetration density is consistent with the relativistic critical density, $\gamma n_c$, which is $\sim 12 ~n_c$ for the initial pulse intensity. The plasma electron density is highly distorted during the interaction. Close to the critical layer, the density is highly modulated at the laser wavelength, and at higher densities it is expelled forming a channel. The electron heat flux in the forward direction is shown to exhibit an annular structure, indicating that, due to the tight focus of the laser, most of fast electrons are accelerated at an angle, which is consistent with experimental observations of annular patterns at the back of a solid target after being hit by an ultraintense laser pulse \cite{bib:norreys}. The plasma electron kinetic energy is maximum at the tip of the laser pulse where considerable electron acceleration occurs.

In conclusion, we have presented 3D simulations of the interaction of ultrahigh intensity laser pulses with overcritical targets. Our results illustrate the importance of a detailed multi-dimensional analysis in order to understand laser absorption and the dynamics of fast electrons, which can be relevant in several scenarios, such as fast ignition of fusion targets and ion acceleration in laser-solid interactions.
           
\section*{Acknowledgment}
 The simulations presented in this paper were performed at the Intrepid supercomputer (ANL) and at the J\"ugene supercomputer (J\"ulich, Germany).


\end{document}